\documentclass{aa} 
 
\input{psfig.tex} 
\def\gsim{\;\lower.6ex\hbox{$\sim$}\kern-7.75pt\raise.65ex\hbox{$>$}\;} 
\def\lsim{\;\lower.6ex\hbox{$\sim$}\kern-7.75pt\raise.65ex\hbox{$<$}\;} 
\begin{document} 
 
%
 
%
\title{ Proton capture elements in the globular cluster NGC 2808 
\thanks{Based on data collected at the European Southern 
Observatory, Chile, during the FLAMES Science Verification}} 
\subtitle{ I. First detection of large variations in Sodium abundances along the 
Red Giant Branch} 
\author{ 
E. Carretta\inst{1}, 
A. Bragaglia\inst{2}, 
C. Cacciari\inst{2} , 
E. Rossetti\inst{2} 
} 
 
\authorrunning{E. Carretta et al.} 
\titlerunning{Proton capture elements in NGC 2808} 
 
 
\offprints{E. Carretta, carretta@pd.astro.it} 
 
\institute{ 
(1) INAF - Osservatorio Astronomico di Padova, vicolo dell'Osservatorio 5, 35122 
 Padova, Italy\\ 
 (2) INAF - Osservatorio Astronomico di Bologna, via Ranzani 1, 40127 Bologna, 
 Italy\\ 
 } 
 
\date{Received ; accepted } 
 
\abstract{ We have used spectra obtained as part of the Science Verification
program of the FLAMES multi-object spectrograph mounted on 
Kueyen (VLT-UT2) to perform an abundance 
analysis of stars along the giant branch (RGB) in the globular cluster NGC 2808. 
Sodium abundances are derived from Na D lines for a sample of 81 cluster stars 
spanning a range of about 2 magnitudes from the tip of the RGB. Our results show 
that a large star-to-star scatter does exist at all positions along the RGB, 
suggesting large variations in the abundance of proton capture elements down to 
luminosities comparable to the red horizontal branch (HB). The distribution of 
Na abundances along the RGB seems to point out that in this  cluster most of 
the observed spread has a primordial origin. Overimposed evolutionary effects, 
if any, must be only at ``noise'' level, at odds with results from a similar 
analysis in M 13 (Pilachowski et al. 1996). This study is a first step  
towards ascertaining if a link exists between the distribution of chemical 
anomalies in light elements along the RGB and the global properties of  
globular clusters, in particular the HB morphology. 
}


\maketitle 
 
\section{INTRODUCTION} 
 
For a long time the globular clusters (GC) have been considered the best 
approximation in the Universe of Simple Stellar Populations (SSP), i.e. 
associations of single, coeval stars sharing the same initial chemical 
composition.
These assumptions have been questioned in the last few years. 
Apart from the relevant presence of binary stars in GC's (showing up as 
Blue Stragglers or eclipsing binaries or peculiar objects, and strongly 
needed by dynamical considerations as a crucial ingredient in the dynamical 
evolution of clusters), also the other two properties that define a 
SSP have been seriously challenged on the basis of modern, accurate studies of 
the chemical composition of cluster stars. 
 
Leaving aside the very peculiar case of $\omega$ Centauri, it has been evident for 
more than 30 years (Osborn 1971) that the hypothesis of monometallicity holds 
for GC only as far as ``heavy'' elements (those belonging to the Fe group) 
are concerned. Early studies (see the comprehensive reviews by Smith 1987 and 
Kraft 1994) based on indexes from photometry or low dispersion spectroscopy 
clearly demonstrated that the lighter elements ({\it in primis} Carbon and 
Nitrogen) showed marked differences along the RGB in several GC's. Moreover, 
striking variations in the CH band strenghts, 
anticorrelated with CN (and NH, when accessible) strenghts were observed in 
several nearby clusters as far down as to main sequence and turn-off stars 
(see e.g. Cannon et al. 1998 and references therein) 
 
Three fundamental steps forward have been made, thanks to the continuous 
progress in the building of efficient spectrographs: 
(i) since the early '90s, the high resolution studies by the 
Lick/Texas group (see Ivans et al. 2001 for updated references) well assessed 
that Na and O abundances are anti-correlated among the first ascent red giant 
branch  (RGB) stars in almost all the clusters surveyed, for at least 1 
magnitude below 
the RGB tip; (ii) the anomalies seem to be restricted to cluster stars, while 
the abundance pattern of field stars is perfectly  explained by the 
classical first dredge-up and a second mixing episode after the level of the 
RGB-bump (Gratton et al. 2000); (iii) the Na-O anticorrelation was recently 
found for the first time among turn-off stars in NGC 6752 by Gratton et al. 
(2001). 
These seminal works point out that not only the CN-cycle (easily detectable 
even with low resolution analysis), but also the ON-cycle of the complete CNO 
H-burning cycle is at work, producing the observed pattern. 
In particular, Na can be produced from $^{22}$Ne by the proton-capture
fusion  mechanism that is at work in the high temperature (inner) regions
where  O is transformed into N (see Denisenkov\& Denisenkova 1990; Langer,
Hoffman \& Sneden 1993).
 
The Na-O anticorrelation is the most convincing evidence of this process that, 
however, is restricted only to high density environments (such as those in the 
globular clusters), since Gratton et al. (2000) showed that Na and O
abundances  are hardly involved in the normal evolution  along   the RGB of
isolated field stars. In a still unknown way, the large number of interactions 
between stars in GCs has to be involved in producing the observed pattern of 
chemical abundances. 

Finally, the recent analysis of high resolution high quality UVES spectra 
of turn-off and early subgiant stars in NGC 6397 and NGC 6752 
(Gratton et al. 2001) and 47 Tuc (Carretta et al. 2003a, in preparation) 
indicated that groups 
of both Na-rich/O-poor and Na-poor/O-rich unevolved stars are present in all 
these clusters. In turn, this must be due to pre-existing abundance variations, 
since these stars do not have the requirements (central high temperatures 
and an extended envelope able to bring  the products of proton 
fusion to the photosphere) to produce the observed chemical pattern as a 
consequence of internal mixing during their evolution. 
 
Primordial variations, viable to explain the observed anomalies, have been 
suggested both in the primordial gas of the proto-cluster (see Cottrell \& Da 
Costa 1981) in the very early phases of star/cluster formation, and as 
Na-rich, O-poor material ejected from a first generation of (now extinct)  
intermediate mass AGB stars that polluted the gas from which a second 
generation of stars (that are presently observed) formed  
(D'antona, Gratton \& Chieffi 1983; Ventura et al. 2001). 
Notice that in this latter case the assumption of strictly coeval 
stars in a GC must be dropped. 
 
On the other hand, the increasingly high N abundances, accompanied by a 
progressive decline in C abundance and $^{12}$C/$^{13}$C ratio, observed at 
increasing luminosity along the RGB in many clusters, testifies that changes 
are going on during the evolution of stars, such as e.g. very deep mixing 
in RGB stars, that may be triggered by enhanced core rotation possibly due 
to the dense cluster environment. 
 
Nowadays, the most likely explanation of the overall chemical pattern observed 
in GC stars is that a contribution of both aspects (primordial and 
evolutionary mixing) is required. The debate seems presently shifted rather 
on how to disentangle these two contributions  and 
to properly ascertain their relative weights within a 
cluster and in clusters of different physical properties (Horizontal
Branch (HB) morphology, 
density, age, metallicity, etc.). 
 
To this purpose large and homogeneous sets of cluster  stars must be observed,
possibly in different evolutionary stages.  This has become possible only
recently, with the advent of efficient   multiobject spectrographs  on large
telescopes, such as  Hydra at the KPNO 4-m and WIYN 3.5 m  telescopes used by
Pilachowski et al. (1996b) and Sneden, Pilachowski \& Kraft  (2000) to survey
the RGB in M13, or  in M15 and M92 respectively. 
 
FLAMES mounted at the 8m Kueyen VLT-UT2 telescope is the ideal instrument 
for this type of studies (see Pasquini et al. 2002 for a technical 
description). 
 
NGC 2808 is by itself a very interesting object, since its HB morphology  (a
red stubby HB and a population of blue HB stars, about 1/4 of the total  HB
stars) has often been labelled as the  second parameter effect working within
an individual cluster.  However, the overwhelming majority of previous works
was  photometric, to investigate the morphology of the color-magnitude diagram
(CMD).  The only spectroscopic study is that of Gratton (1982), who analyzed a
single  giant star obtaining a metallicity of [Fe/H]$=-1.1$ dex\footnote{We
use the usual  spectroscopic notation: log~n(A) is the abundance (by  number)
of the element A in the usual scale where log~n(H)=12; notation [A/H] is  the
logarithmic ratio of the abundances of elements A and H in the star, minus  the
same quantity in the Sun.}. 
 
This situation dramatically changed at the end of January 2003, when $\sim$ 130 
RGB stars in NGC 2808 were observed during the program "Mass loss in Globular 
Cluster Red Giant Stars" (proposed by C. Cacciari and A. Bragaglia), that was
part of the FLAMES Science Verification program. The results of this study 
are presented in a separate paper (Cacciari et al. 2003, in preparation).  
In the present paper we exploit part of this material  for the analysis of 
the Sodium D lines in order to get insights on the distribution of this proton 
capture element along the RGB. 
 
In the next section we present the star selection (that was performed for 
studying the mass loss process and was not optimized for the present purpose) 
and the observations. In Sect. 3 and 4, respectively, we  explain the 
atmospheric parameters adopted and the abundance analysis for Na, whose 
results are discussed in Sect. 5. Comparison with results for other 
clusters are made in Sect. 6, and a short summary is given in Sect. 7.

\begin{figure} 
\psfig{figure=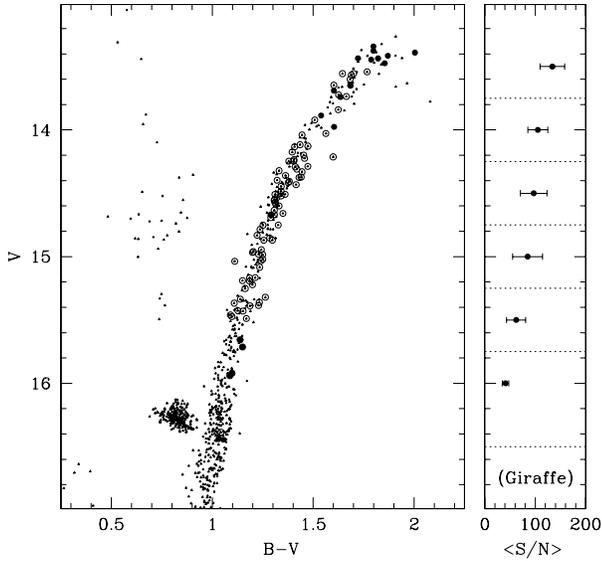,width=8.8cm,clip=} 
\caption[]{Left panel: $V$, $B-V$ colour-magnitude diagram of NGC 2808 
(Piotto  et al. 2003, in preparation). Stars observed with the GIRAFFE HR 12
grating in the  Medusa  configuration are indicated by open circles, while
filled circles are  stars observed with fibers feeding the UVES Red Arm. Right
panel: average  S/N in 0.5 mag bins along the RGB.} 
\label{f:observedstar} 
\end{figure}

\section{SAMPLE SELECTION AND OBSERVATIONS} 
 
Our program stars were observed in 2003, January 24-25 during 
the FLAMES Science Verification program at the ESO Paranal Observatory. 
The data were released to the public on 2003, March 3. 
The night was not photometric, the sky conditions were thin cirrus and the 
seeing was about 0.9 arcsec.   
86 stars were observed with the grating HR 12 of GIRAFFE, 
giving an useful spectral range from 5820 to 6134~\AA, with a resolution of 
about 15,000 at the centre of spectra. During all GIRAFFE exposures, the 
available fibers (8 for each setup, including one fiber on the sky) feeding 
the Red Arm of the high resolution spectrograph UVES were centered on other 
RGB stars, for a total of 20 objects observed at high resolution 
(R$\sim 45,000$). Only 4 stars were observed by both GIRAFFE and UVES. 
Details of these UVES observations and their analysis are 
deferred to a forthcoming paper (Carretta et al. 2003b, in preparation). 
Here, we concentrate only on the GIRAFFE sample in the Na D region. 
 
Fig.~\ref{f:observedstar} indicates the position of our 
program stars in the $V$, $B-V$ color-magnitude diagram of NGC 2808. 
We remind here that the main purpose 
of this program was to study the mass loss along the RGB, so the observations 
are concentrated in the region close to the RGB tip. 
Cool and (moderately) metal-rich stars are not the best targets for abundance 
analysis due to the line crowding, affecting the continuum positioning, and to 
concerns on existing model atmospheres that are unable to well 
reproduce the cooler red giants. 
 
\begin{figure} 
\psfig{figure=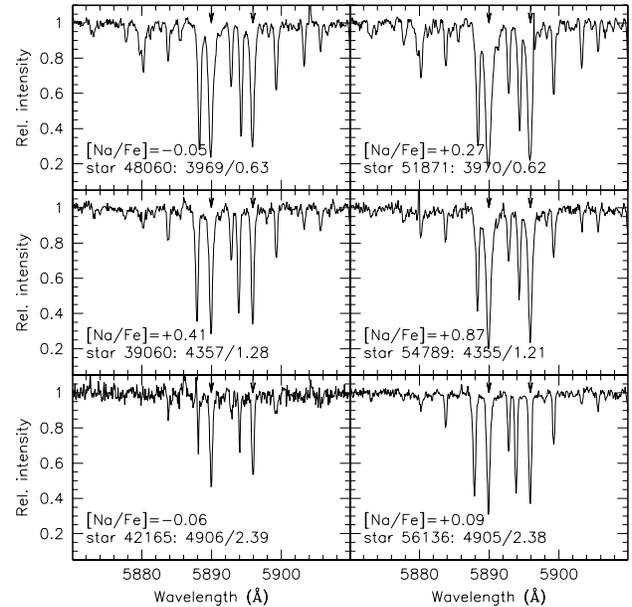,width=8.8cm,clip=} 
\caption[]{ 
Typical spectra for bright (upper panels), intermediate (middle panels) 
and faint (lower panels) objects in our sample with different Na abundances. 
 The small arrows 
indicate the position of the stellar Na I D lines; notice how well they 
can be separated from the interstellar lines, here shifted to the blue. 
We also indicate for each star the values of T$_{\rm eff}$, $\log\, g$, 
and [Na/Fe]. 
S/N (per pixel) varies from about 100, to 85, to 45 for the three 
magnitude levels.} 
\label{f:tipicisp} 
\end{figure} 
 
In Fig.~\ref{f:tipicisp} typical spectra are displayed for stars spanning all 
the range of parameters sampled along the RGB. 
 
The sample goes from the tip of the RGB (about $M_V = -3$, adopting a distance 
modulus (m-M)$_V=$15.59 from Harris 1996) to just below the level of the red HB 
(about $M_V = 1$). Only two stars have been observed below $V=16$, the 
bulk of program stars lie within 2 mag from the RGB tip. 
The right panel in Fig.~\ref{f:observedstar} shows how the average S/N ratio 
varies as a function of the V magnitude along the RGB. The scatter around a 
mean value at fixed magnitude is likely due to a combination of small 
mis-positioning errors of the star in the individual fibers and different 
throughtputs in transmission from different fibers.

\begin{figure} 
\psfig{figure=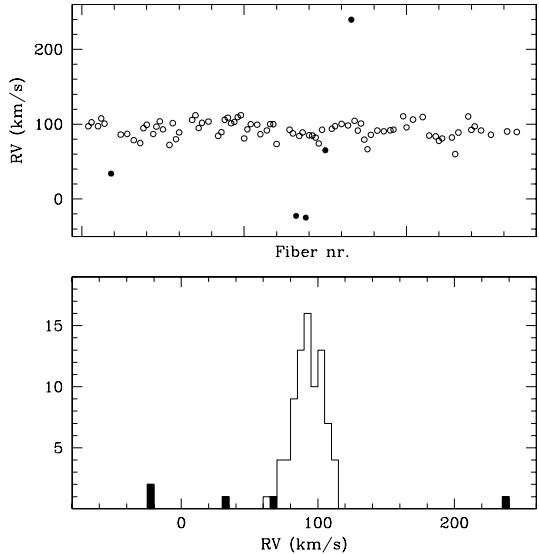,width=8.8cm,clip=} 
\caption[]{Upper panel: measured radial velocities for program stars plotted 
as a function of the fiber numbers. Filled circles are stars rejected from 
our analysis as non-members. Lower panel: histogram of the 
radial velocities, where the rejected stars are indicated by the filled 
histogram.} 
\label{f:istovr} 
\end{figure}

Cluster membership was mostly secured by choosing stars near the mean ridge 
line of 
the giant branch and checked {\it a posteriori} by measuring radial velocities 
from spectral lines. 
The radial velocity (RV) for each star was derived by fitting Gaussian profiles 
to about 30 lines (with the IRAF\footnote{  
IRAF is distributed by the NOAO, which are operated by AURA, under contract  
with NSF}    
routine $rvidlines$). 
 
Fig.~\ref{f:istovr} shows the distribution of radial 
velocities. A small zero point offset between fibers is possible, due to the 
reduction procedure not optimized for precise radial velocities measurements, 
that could reach $\pm$ 10 kms$^{-1}$, according to the ESO relevant web 
pages. We have not done any detailed check on radial velocities, 
since it was outside our main 
purpose; we only note that the 4 stars that were observed also with UVES 
at high resolution - and more reliable wavelength calibration - 
show a difference between derived velocities from 1 to --8 kms$^{-1}$, 
with an average of --3.5 kms$^{-1}$, 
and measurement of the telluric Na I line at 5889.884 \AA 
~in  star \#56136 indicates a velocity zero point of 3.5 kms$^{-1}$. 
So the derived RV's are in general precise enough to discriminate 
between cluster and field stars. 
Using the RV distribution, we disregarded 4  stars as sure 
field interlopers. Another star was rejected from the sample after the 
abundance analysis (see below, Sect. 4). 
The average heliocentric velocity is RV = 102.1 kms$^{-1}$ 
($\sigma= 10.9$), after eliminating these 5 non-members. 
 
The final sample consists therefore in 81 red giant stars. 
What is the physical evolutionary status of these giants? Are they truly 
first ascent giant stars or is the sample contaminated by stars in a more 
evolved stage, i.e. Asymptotic Giant Branch (AGB) stars? 
For the brightest objects there is no way to discriminate between RGB and 
AGB stars either photometrically or spectroscopically: 
their position in the CMD is almost identical, and small 
differences in colours may well be due to photometric errors and/or 
variability, and in both cases emissions on the H$\alpha$ wings can be 
present. Anyway, from robust evolutionary theory expectations, AGB 
stars should represent only a small fraction ($\sim$ 10\%) of the 
brightest objects. 
When we consider lower luminosity stars, photometric distinction is easier, 
and we may also hope to detect AGB stars via  H$\alpha$ emission 
wings (as done by Sneden et al. 2000): this emission, which is rather common 
also in RGB stars near the tip (see e.g. Lyons et al. 1996, and the vast 
literature cited there), is not present in fainter objects. 
Hence, if stars with M$_V \lsim$ --1 or fainter  present H$\alpha$ 
emission wings, they may be taken as -- or at least suspected to be -- 
AGB stars. 
Cacciari et al. (2003, in preparation) have done a similar survey of the 
same stars we are analysing here, and detected emissions only for one such 
star (\#10341): we will still retain it in our sample, cautioning that it
could well not be a true RGB star.

\section{ATMOSPHERIC PARAMETERS} 
 
As described in Cacciari et al. (2003), the effective temperatures 
T$_{\rm eff}$ and bolometric corrections B.C. for our stars were 
derived from the $V-K$ colors in the 2MASS All-Sky Data Release 
(accessible at {\it www.ipac.caltech.edu/2mass/releases/allsky/} 
and released on March 25, 2003), and using the calibrating 
equations by Alonso et al. (1999, with the erratum of Alonso et 
al. 2001). We have adopted  A$_K=0.353E(B-V)$ and A$_V=3.1E(B-V)$ 
from Cardelli et al. (1989), and $E(B-V)$=0.22 from Harris (1996). 
 
The 2MASS photometry was transformed to the TCS system and an 
input metallicity of [Fe/H]$=-1.25$ was adopted\footnote{This 
value is intermediate among several previous determinations 
(cf. Walker 1999), 
and is slightly different from the metallicity we adopt in the present 
analysis; however, the dependence of ($V-K$) on [Fe/H] is so weak that 
temperatures are almost unaffected by this difference}. 
 
Surface gravities $\log\, g$ were obtained from effective temperatures and 
bolometric corrections, using the distance modulus $(m-M)_V=15.59$, and 
assuming that the stars have masses of 0.85 M$_\odot$. The adopted 
bolometric magnitude of the Sun is $M(bol)_\odot = 4.75$. 
 
For the overall metallicity, we adopted the average metal abundance as 
derived by the analysis of the high resolution UVES red spectra (Carretta et 
al. 2003b, in preparation): 
[A/H]$=-1.14 \pm 0.01$ dex, $\sigma=0.06$ dex (19 
stars)\footnote{This in turn is based on a solar Iron abundance of 
logn(Fe I)$_\odot = 7.54$, see Gratton et al. (2001).}. This 
choice stems from the small number of Fe I lines  available in our spectral 
range (about 17 lines, on average, are measurable per star) and from the 
rather low resolution. 
As discussed, e.g., in Carretta \& Gratton (1997), at lower resolution 
line blends are more frequent, leading to overestimate  the equivalent widths 
(EW) in a complex way that is a function of metallicity, S/N and 
temperature (all affecting the line crowding), as well as resolution. 
Since the RGB magnitude range monitored by GIRAFFE observations 
was covered also by UVES observations, we believe that the use of UVES-based  
metallicities is justified and does not bias the results we obtain from 
GIRAFFE data. 
 
Finally, based on the same considerations, we adopted the microturbulent 
velocity $v_t$ derived from  the high resolution UVES spectra, using a large 
number of Fe I lines of different strenghts. The resulting relation is well 
constrained, so we adopted $v_t$=1.73 kms$^{-1}$ for T$_{\rm eff} < 4100$ K 
and 
$v_t = -1.3 \cdot 10^{-4} \times T_{\rm eff} +7.10$ for T$_{\rm eff} \geq 4100$. 
 
Star designations and derived atmospheric parameters are listed in the first 
four column of Table~\ref{t:parabu}. 
 
\begin{table} 
\caption{Atmospheric parameters and derived Na abundances, both in LTE and
in NLTE. 
Star designations are taken from Piotto et al. (2003)} 
\begin{tabular}{lcccccc} 
\hline 
Star & T$_{\rm eff}$ & $\log g$ & $v_t$ & [Na/Fe] &
 [Na/Fe] & $\sigma$ \\ 
&&& &LTE & NLTE & \\
     & (K) & dex & kms$^{-1}$    &dex    & dex  & dex \\ 
\hline 
\hline 
7536  &  4245 &  1.18 & 1.58 &$-$0.16  &  +0.06 & 0.04 \\
7558  &  4691 &  1.87 & 1.00 &  +0.17  &  +0.28 & 0.09 \\
7788  &  4459 &  1.53 & 1.30 &  +0.12  &  +0.29 & 0.15 \\
8603  &  4290 &  1.24 & 1.52 &$-$0.24  &$-$0.02 & 0.13 \\
8739  &  4211 &  1.12 & 1.63 &$-$0.08  &  +0.14 & 0.15 \\
8826  &  4564 &  1.66 & 1.17 &  +0.42  &  +0.57 & 0.16 \\
9230  &  4132 &  0.96 & 1.73 &  +0.04  &  +0.29 & 0.02 \\
10012 &  4243 &  1.12 & 1.58 &  +0.23  &  +0.45 & 0.04 \\
10105 &  4455 &  1.44 & 1.31 &$-$0.05  &  +0.13 & 0.01 \\
10201 &  4717 &  2.02 & 0.97 &  +0.42  &  +0.52 & 0.08 \\
10265 &  4283 &  1.18 & 1.53 &  +0.24  &  +0.45 & 0.04 \\
10341 &  4399 &  1.40 & 1.38 &  +0.44  &  +0.63 & 0.04 \\
10571 &  4252 &  1.19 & 1.57 &$-$0.01  &  +0.21 & 0.08 \\
13575 &  4569 &  1.80 & 1.16 &$-$0.23  &$-$0.10 & 0.01 \\
13983 &  4826 &  2.17 & 0.83 &  +0.35  &  +0.42 & 0.16 \\
30523 &  4733 &  1.94 & 0.95 &  +0.26  &  +0.37 & 0.05 \\
30900 &  4666 &  1.74 & 1.03 &  +0.21  &  +0.34 & 0.11 \\
31851 &  4731 &  1.83 & 0.95 &  +0.77  &  +0.89 & 0.03 \\
32469 &  4379 &  1.14 & 1.41 &  +0.44  &  +0.65 & 0.11 \\
32685 &  4788 &  2.03 & 0.88 &  +0.37  &  +0.47 & 0.01 \\
32862 &  4525 &  1.69 & 1.22 &  +0.15  &  +0.30 & 0.10 \\
32909 &  4572 &  1.66 & 1.16 &  +0.29  &  +0.44 & 0.09 \\
32924 &  4538 &  1.57 & 1.20 &  +0.54  &  +0.70 & 0.15 \\
33918 &  4312 &  1.21 & 1.49 &  +0.06  &  +0.27 & 0.09 \\
35265 &  4547 &  1.76 & 1.19 &  +0.07  &  +0.21 & 0.10 \\
37496 &  4547 &  1.58 & 1.19 &  +0.26  &  +0.42 & 0.03 \\
37998 &  4317 &  1.18 & 1.49 &  +0.12  &  +0.34 & 0.08 \\
38228 &  4588 &  1.65 & 1.14 &  +0.38  &  +0.52 & 0.01 \\
38244 &  4659 &  1.78 & 1.04 &  +0.23  &  +0.36 & 0.04 \\
38967 &  4717 &  1.83 & 0.97 &  +0.41  &  +0.53 & 0.08 \\
39060 &  4357 &  1.28 & 1.44 &  +0.21  &  +0.41 & 0.08 \\
40983 &  4364 &  1.26 & 1.43 &  +0.29  &  +0.49 & 0.01 \\
41008 &  4684 &  1.79 & 1.01 &  +0.67  &  +0.80 & 0.03 \\
41828 &  4409 &  1.36 & 1.37 &  +0.11  &  +0.30 & 0.22 \\
41969 &  4377 &  1.30 & 1.41 &  +0.07  &  +0.26 & 0.01 \\
42073 &  4268 &  1.17 & 1.55 &$-$0.02  &  +0.20 & 0.13 \\
42165 &  4906 &  2.39 & 0.72 &$-$0.09  &$-$0.06 & 0.04 \\
42789 &  4813 &  1.93 & 0.84 &  +0.78  &  +0.89 & 0.09 \\
42886 &  4791 &  2.14 & 0.87 &  +0.11  &  +0.19 & 0.08 \\
42996 &  4769 &  1.93 & 0.90 &  +0.29  &  +0.39 & 0.09 \\
43041 &  4035 &  0.70 & 1.73 &$-$0.23  &  +0.05 & 0.04 \\
43247 &  4463 &  1.50 & 1.30 &  +0.29  &  +0.46 & 0.09 \\
43794 &  4431 &  1.38 & 1.34 &  +0.05  &  +0.24 & 0.13 \\
43822 &  4411 &  1.39 & 1.37 &  +0.16  &  +0.34 & 0.03 \\
44665 &  4310 &  1.18 & 1.50 &  +0.17  &  +0.39 & 0.16 \\
44984 &  4405 &  1.36 & 1.37 &$-$0.00  &  +0.19 & 0.13 \\
45443 &  4425 &  1.39 & 1.35 &  +0.05  &  +0.23 & 0.13 \\
45840 &  4220 &  1.06 & 1.61 &$-$0.14  &  +0.10 & 0.03 \\
46041 &  4568 &  1.63 & 1.16 &  +0.26  &  +0.41 & 0.04 \\
46367 &  4383 &  1.46 & 1.40 &  +0.21  &  +0.38 & 0.03 \\
46663 &  4474 &  1.41 & 1.28 &  +0.47  &  +0.65 & 0.04 \\
46868 &  4463 &  1.50 & 1.30 &  +0.05  &  +0.22 & 0.02 \\
47145 &  4051 &  0.74 & 1.73 &  +0.02  &  +0.29 & 0.05 \\
47421 &  4288 &  1.14 & 1.53 &  +0.21  &  +0.43 & 0.11 \\
48011 &  4375 &  1.28 & 1.41 &  +0.63  &  +0.83 & 0.01 \\
48060 &  3969 &  0.63 & 1.73 &$-$0.35  &$-$0.05 & 0.04 \\
48128 &  4368 &  1.31 & 1.42 &  +0.48  &  +0.68 & 0.08 \\
\hline 
\end{tabular} 
\label{t:parabu} 
\end{table} 
 
\addtocounter{table}{-1} 
 
\begin{table} 
\caption{(continue)} 
\begin{tabular}{lcccccc} 
\hline 
Star & T$_{\rm eff}$ & $\log g$ & $v_t$ & [Na/Fe] &
 [Na/Fe] & $\sigma$ \\ 
&&& &LTE & NLTE & \\
     & (K) & dex & kms$^{-1}$    &dex    & dex  & dex \\ 
\hline 
\hline 
49509 &  4164 &  0.94 & 1.69 &  +0.12  &  +0.37 & 0.08 \\
49680 &  3952 &  0.62 & 1.73 &$-$0.37  &$-$0.07 & 0.05 \\
49743 &  4665 &  1.77 & 1.04 &  +0.46  &  +0.60 & 0.06 \\
49942 &  4023 &  0.75 & 1.73 &$-$0.15  &  +0.13 & 0.02 \\
50371 &  4054 &  0.82 & 1.73 &$-$0.23  &  +0.04 & 0.07 \\
50861 &  4039 &  0.69 & 1.73 &  +0.13  &  +0.41 & 0.06 \\
50866 &  4432 &  1.42 & 1.34 &  +0.40  &  +0.58 & 0.09 \\
50910 &  4577 &  1.64 & 1.15 &  +0.25  &  +0.40 & 0.02 \\
51416 &  4546 &  1.61 & 1.19 &  +0.39  &  +0.54 & 0.09 \\
51515 &  4154 &  1.05 & 1.70 &  +0.04  &  +0.28 & 0.06 \\
51646 &  4886 &  2.01 & 0.75 &  +0.88  &  +0.98 & 0.08 \\
51871 &  3970 &  0.62 & 1.73 &$-$0.02  &  +0.27 & 0.10 \\
51963 &  4242 &  1.11 & 1.59 &  +0.06  &  +0.28 & 0.04 \\
52006 &  4397 &  1.27 & 1.38 &  +0.62  &  +0.81 & 0.09 \\
54264 &  4668 &  1.84 & 1.03 &  +0.15  &  +0.27 & 0.09 \\
54284 &  4327 &  1.26 & 1.47 &  +0.14  &  +0.34 & 0.11 \\
54733 &  4236 &  1.08 & 1.59 &  +0.25  &  +0.48 & 0.05 \\
54789 &  4355 &  1.21 & 1.44 &  +0.66  &  +0.87 & 0.01 \\
55031 &  3995 &  0.67 & 1.73 &$-$0.35  &$-$0.05 & 0.05 \\
55437 &  4440 &  1.44 & 1.33 &  +0.27  &  +0.45 & 0.05 \\
55609 &  4430 &  1.46 & 1.34 &  +0.23  &  +0.41 & 0.04 \\
55627 &  4459 &  1.41 & 1.30 &  +0.52  &  +0.71 & 0.13 \\
56136 &  4905 &  2.38 & 0.72 &  +0.04  &  +0.09 & 0.09 \\
56710 &  4643 &  1.86 & 1.06 &  +0.04  &  +0.16 & 0.09 \\
\hline 
\end{tabular} 
\label{t:parabu} 
\end{table}

\section{ANALYSIS: SODIUM ABUNDANCES} 
 
Equivalent width (EW) of Na D lines were measured following the method outlined 
in Bragaglia et al. (2001): we refer to that paper for further details. The 
fraction of the (highest) spectrum points to be used by the automatic program in 
order to set a local continuum centered on the feature of interest is the 
crucial point of the procedure. 
We checked this parameter by inspecting by eye the continuum location 
as defined by the program for 
some lines of typical spectra distributed all over the 
temperature range. After this test, we set this fraction to 1/3 for stars with 
T$_{\rm eff} \leq 4200$ K and to 1/2 for warmer stars. 
 
Using this value and the atmospheric parameters derived in the 
previous section, we 
measured the EWs of the Na D lines at 5889.97 and 5894.94 \AA ~for all 
81 program stars, and derived the Na abundances interpolating in the 
Kurucz (1995) grid of model atmospheres with the overshooting option set on. 
 
However, a Gaussian function is a poor approximation for fitting these very 
strong lines and may give misleading results, missing the relevant 
contribution of the damping wings of the lines. 
For these reasons we decided to adopt the Na abundances obtained from 
comparison with synthetic D1 and D2 lines: the procedure is much more 
time consuming but gives more reliable results than a Gaussian fit or 
even a direct integration. Moreover, concerns related to the positioning 
of the continuum are much reduced. 
 
Hence, we computed a grid of synthetic spectra for each of the Na D lines. 
Each grid includes 21 spectra with T$_{\rm eff}$ ranging from 3950 to 4950 K 
(this range brackets all our program stars) in steps of 50 K. 
Using the single value [A/H]$=-1.14$ dex, $v_t$ for each synthetic spectrum 
was computed using the relation obtained from the UVES spectra, 
as described in the previous section;  the gravity was derived from a 
T$_{\rm eff}$-$\log\, g$ relation obtained with a linear fit from our 
program stars. 
 
For each of the 21 spectra (i.e. values of T$_{\rm eff}$) in our grid  
we computed 10 synthetic spectra of the Na line by varying the [Na/Fe] 
ratio from --0.4 to +1.4 dex, in steps of 0.2 dex. 
Thus we could associate to each Na line a set of 210 
synthetic spectra spanning the relevant range both in temperature and Na 
abundances. These synthetic spectra were convolved with a Gaussian function to 
match the resolution of the GIRAFFE spectra. We found that a smoothing FWHM of 
0.3 \AA ~is appropriate for our program stars. 
 
Starting with input Na abundances determined by line analysis, we chose the 
nearest value of temperature for each star, and compared the observed 
spectrum to the 3 synthetic spectra bracketing the Na abundance from EW. 
Visual inspection was used to obtain the Na abundances from each line. 
For the star Na abundance we then took simply the average from the two Na D 
lines, adopting log n(Na)=6.23 as the solar Na abundance. 
 
\begin{figure} 
\psfig{figure=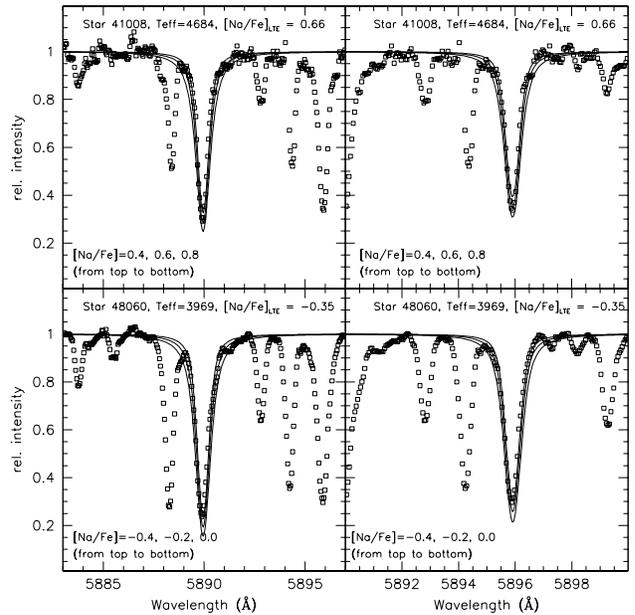,width=8.8cm,clip=} 
\caption[]{Upper left panel: spectrum synthesis of the Na D2 line at 
5889.97~\AA\ in star 41008. 
Open squares are the observed spectrum, while lines represent the synthetic 
spectra computed at 4700 K for 3 different Na abundances: [Na/Fe]=0.4, 0.6 
and 0.8 dex from top to bottom, respectively. Upper right panel: the same, 
but for the Na D1 line at 5895.94~\AA. Lower left panel: spectrum synthesis 
of the Na D2 line in star 48060, near the RGB tip. The plotted 
synthetic spectra are computed at 3950 K and for abundance ratios of 
[Na/Fe]= -0.4, -0.2 and 0.0 from top to bottom. Lower right panel: the same, 
but for the Na D1 line. All synthetic spectra have been convolved with a 
Gaussian having FWHM=0.3~\AA\ to take into account the instrumental profile 
of the observed spectra. All abundances are given in the LTE assumption.} 
\label{f:sintesina} 
\end{figure}

In Fig.~\ref{f:sintesina} we show two examples of the fitting of synthetic 
spectra to the observed Na D lines for two stars in different position 
along the RGB. In the upper panels, the star 41008 
has T$_{\rm eff} = 4684$ K and was compared to spectra computed at 4700 K, 
while in the lower panels the program star has T$_{\rm eff}= 3969$ and 
Na abundances 
are obtained through comparison with synthetic spectra computed at 3950 K. 
 
This Figure shows that despite the difference of $\sim 1$ dex in Na abundance 
and $\sim$ 800 K in temperature, the Na features are well reproduced. 
 
A comparison of Na abundance derived by line analysis (EWs) and by spectrum 
synthesis is highly instructive. The average abundance difference (in the 
sense synthesis minus line) is $\Delta$[Na/Fe]=+0.30$\pm$0.01 dex, with an 
r.m.s.=0.09 dex (81 stars\footnote{ 
We exclude from the sample and from the following discussion star 37781, 
that is probably not-member of the cluster  even if its radial velocity 
is compatible with membership. The unusually high Na abundance obtained for 
this stars ([Na/Fe]$\simeq +1.5$) is likely due to an erroneous estimate of 
atmospheric parameters due to the initial assumption that this star is member 
of NGC 2808, or because it might be an AGB star.}). 
This test illustrates how  severely the Gaussian fitting 
approximation may  underestimate the measure of  EWs for these 
strong lines, resulting in spurious Na abundances. 

Since the observations were made for other purposes and did not aim at 
deriving chemical abundances, rapidly rotating early type stars were not 
observed. Lacking a suitable template to subtract telluric lines, we 
evaluated their impact on derived abundances by estimating their contribution 
to the EW of the Na lines within the region where the fitting of synthetic 
spectra was performed. We verified that the possible effect of telluric lines 
on the derived abundances amounts to no more than a few hundredths of a dex 
and was therefore considered negligible. 
 
Finally, we neglected the hyperfine splitting for Na D lines. Explorative 
computations with atomic parameters for the components of each line taken from 
McWilliam et al. (1995) allowed us to conclude that the effect on our Na 
abundances is negligible, likely because in very strong lines the effect 
itself is reduced since all hyperfine components in the line core are saturated 
(see McWilliam et al. 1995).

\subsection{NLTE corrections} 
 
Statistical equilibrium computations (e.g. Bruls et al. 1992, Gratton et al. 
1999) show that the NLTE abundance corrections for Na D lines follow 
a rather complicated pattern, due to the complex relative interplay between  
photoionization and the so-called photo suction effect (Bruls et al. 1992) 
that consists in a recombination cascade to the ground level, where the D 
lines originate. 
Moreover, the strong Na D lines form high in the atmosphere, where 
departures from LTE are larger. 
In order to assess the true Na distribution, a careful evaluation of these  
effects is then required. 
  
We derived appropriate corrections for departures from the LTE following the 
prescriptions of Gratton et al. (1999). Since these corrections are a function 
of both stellar physical status (temperature and gravity) and line strength, 
we measured the equivalent widths of synthetic Na D2 and D1 lines for 
all (T$_{\rm eff}$, [Na/Fe]) pairs. These EWs were used to construct 
interpolating relations (one for each line) as a function of temperature 
and Na abundance. 
 
Next, we entered in these relations with stellar temperatures and previously 
derived LTE 
Na abundances from spectrum synthesis for all our program stars, deriving 
[Na/Fe] ratios properly corrected for departures from LTE following Gratton et 
al. (1999). These corrected abundances are listed in Table~\ref{t:parabu} 
(labeled as [Na/Fe]$_{NLTE}$) and represent our final set of abundances on which 
the following considerations and discussions are based.  As reference for
the reader, in this Table are also listed abundances derived under the LTE
assumption. 
The errors listed in Table~\ref{t:parabu} are the r.m.s. of the straight average 
between the abundances as given by Na D2 and D1 lines. 
 
\begin{figure} 
\psfig{figure=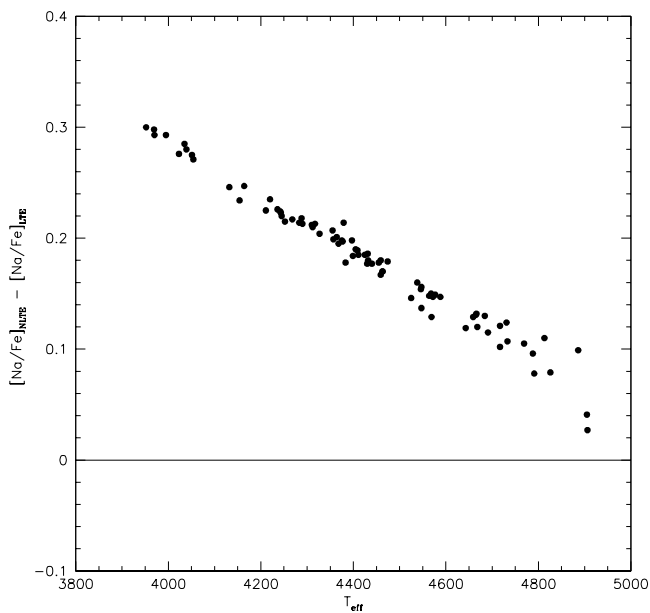,width=8.8cm,clip=} 
\caption[]{Corrections to the Na abundances for departure from LTE as a 
function of effective temperature, derived for our program stars.} 
\label{f:corrnlte} 
\end{figure}

In Fig.~\ref{f:corrnlte} we show the NLTE corrections to [Na/Fe] LTE 
abundances as a function of effective temperature.   As one can see, the 
corrections are progressively larger going from lower luminosity giants toward 
giants near the tip of the RGB (where they can reach as much as 0.3~dex). 
Neglecting these corrections would then result in a spurious trend as a 
function of T$_{\rm eff}$, with tip giants having 0.2-0.3 dex lower 
[Na/Fe] abundances than less evolved RGB stars.

\section{RESULTS AND DISCUSSION} 
 
\begin{figure} 
\psfig{figure=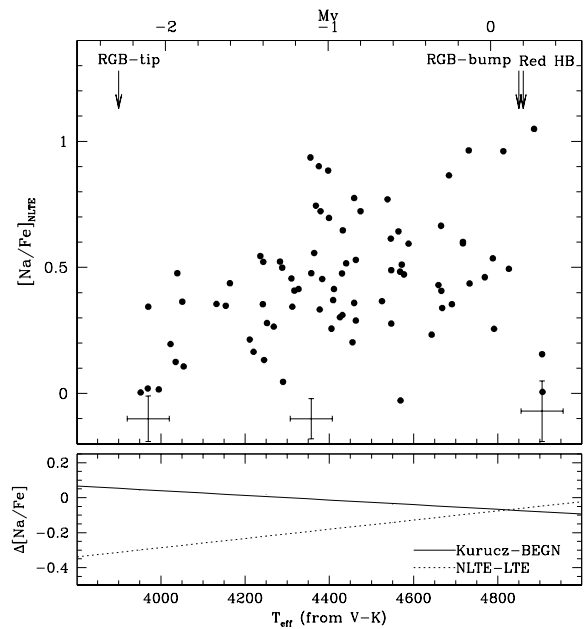,width=8.8cm,clip=} 
\caption[]{Upper panel: abundance ratios [Na/Fe] for stars along 
the RGB of NGC 2808, 
corrected for the effects of departures from LTE, as a function of the effective 
temperatures. At the top of the figure the approximate 
luminosity levels for the RGB-tip, the RGB-bump and the average luminosity 
of the red extreme of the HB are indicated. At 
the bottom of the figure we show the typical error bars over the whole range 
of temperatures spanned by our sample, due to random uncertainties in the 
adopted atmospheric parameters and measurement errors. \\
Bottom panel: the solid line indicates the difference in [Na/Fe] derived 
by using Bell et al. (1976) model atmospheres with respect to Kurucz 
(1995) models. 
The dashed line indicates the effect of NLTE vs LTE approximations on the 
derived [Na/Fe] abundances.} 
\label{f:nateff} 
\end{figure}

Our final [Na/Fe] abundances (including NLTE corrections) are plotted as a 
function of temperature in Fig.~\ref{f:nateff}. 
 
The first clear evidence that can be derived from this plot is the large 
star-to-star scatter, at every luminosity/temperature along the RGB.  
Also, the average Na abundance is generally quite high. Neither 
feature is surprising in itself, as both have been commonly found in several 
other clusters; now they are found for the first time also in 
this cluster, that had never been studied in detail from a spectroscopic point 
of view before. 
 
When compared to abundances of field stars of similar metallicity (e.g. Gratton 
et al. 2000, Pilachowski et al. 1996a), these results confirm once more that 
globular cluster stars present a pattern of abundances suggestive of processes 
involving elements produced somewhere deep in the star, near the H-burning 
shell, by proton-capture nucleosynthesis that operates at high temperatures, 
and likely producing Na through the NeNa cycle. 
 
In fact, while field stars have generally low Na abundances (almost coincident 
with the lower envelope of the cluster abundances), with small scatter 
around the average value (comparable with observational errors), in cluster 
stars the standard evolutionary paradigm is broken. 
The spread in Na abundances in NGC 2808 reaches as much as 1 dex and does  
not decrease with temperature much below 0.5 dex, in [Na/Fe]. 
 
Apart from the well known Na-O anticorrelation (also present among RGB stars
of  NGC 2808 at a level that could be at least comparable with that
existing in M 13,  Carretta et al. 2003b), our results from Na abundances alone
allow us  to conclude that the standard first dredge-up and an additional
mixing  mechanism at luminosities above the RGB-bump are not enough to explain
the  observed pattern, at odds with what happens in field stars. 
 
Na abundances so large and scattered from star to star do require that a 
variable amount of Na-enriched matter polluted the stars in the past, or 
that the efficiency of some further (deep) mixing mechanism varies along 
the RGB modifying pre-existing stellar abundances.  
Thus, the question is now: is the observed scatter a cosmic scatter 
intrinsic to the studied sample, or is it an artifact of the analysis 
due to conjuring effects of uncertainties on derived abundances? 
 
In general, we may consider  three sources of errors on abundance 
determinations, some of which may affect the scatter, and some others 
only the zero point. 
First, the adopted temperature or oscillator strength scales may be 
wrong\footnote{Note however that a systematic error of about 100 K in the 
temperature scale would give a difference of about 0.2 dex in abundances from 
neutral and singly ionized Fe lines, while the average difference found by 
Carretta et al. (2003b, from whom we adopted the average metallicity for NGC 
2808) is only $-0.01$ dex, implying a rather small systematic error on the 
temperatures.
Moreover, the atomic parameters for the strong Na D lines are well known, as 
well as the laboratory oscillator strengths for many of the Fe lines used to 
derive the abundances from the UVES spectra.}. 
However, this would result only in a zero point shift, affecting rigidly 
the entire metallicity distribution but leaving almost unchanged the 
star-to-star spread. 
 
Second, other effects exist, that may result in spurious systematic trends 
on the results if not taken into account. A clear example is provided by 
the corrections for departure from LTE considered in Sect. 4.1. In 
the bottom panel of 
Fig.~\ref{f:nateff} we have shown (dashed line) the 
changes in [Na/Fe] abundances that would result by neglecting the non-LTE 
corrections. 
Since, as shown by Fig.~\ref{f:corrnlte} and discussed in Gratton 
et al. (1999), 
the amount of the correction increases at lower temperatures and 
gravities, had we neglected this effect a marked trend would appear in the 
results, with Na abundances decreasing as the stars approach the RGB-tip. 
Notice that this effect should be in principle larger at low metallicity. 
This should be borne in mind when comparing the pattern found in different 
clusters, such as the survey of Na abundances in giants of M 15 and M 92 by 
Sneden et al. (2000) where Na D lines were also used, but no correction 
for departures from LTE was applied. 
 
Another effect of some impact may be the choice of the grid of atmospheric 
models used in the analysis. In the present study we adopted the Kurucz (1995; 
K95) models with the overshooting option. We then repeated the analysis by 
computing the Na abundances using the Bell et al. (1976; hereinafter BEGN) 
model atmospheres, that are maybe a bit better at reproducing the atmospheres 
of cooler stars such as those at the RGB-tip. 
Fig.~\ref{f:nateff} shows (bottom panel solid line) that at T$_{\rm eff}$ 
about  3900 K the BEGN models provide [Na/Fe] ratios about 0.04 dex larger than 
those derived with K95 models. At lower luminosity, around T$_{\rm eff} \sim 
4900$ K,  the trend is inverted, with abundances from BEGN models being about 
0.08 dex lower than those from K95. 
The net effect is again a (small) trend; if one used BEGN models and no 
corrections for non-LTE, the two opposite trends would  nearly compensate 
each other. Anyway, also this effect would 
only tilt the distribution function of Na abundances along the RGB, leaving 
the rms scatter untouched. 
 
Finally, random errors due to uncertainties in the adopted atmospheric 
parameters and to measurement errors may potentially add a random noise to the 
observed abundances. Are they enough to explain the observed scatter? 
In order to test this, we evaluated the sensitivity of abundances to errors 
in the adopted atmospheric parameters in the usual way, i.e.  
we repeated the analysis changing by a given amount one parameter at a time 
(T$_{\rm eff}$, $\log\, g$, [A/H] and $v_t$). 
We performed this test on three typical stars spaced all along the 
RGB, the results of this exercise are summarized in Table~\ref{t:sensitivity}. 
 
To correctly use this Table, one has to notice that the changes in the 
parameters are arbitrary,  and have been adopted only for demonstrative 
purposes.  For example, 
realistic errors in T$_{\rm eff}$ are not likely to exceed $\sim 50$ K, 
whereas in column 2 of Table~\ref{t:sensitivity} we have used twice 
this value.  Similarly, the gravity is unlikely to be wrong by much more 
than 0.1 dex and a conservative estimate of errors in metallicity is about 
0.1 dex (Carretta et al. 2003b), hence the values listed  in column 3 and 4 
should be scaled accordingly.  
Moreover, since our derivation of atmospheric parameters uses the position of 
stars in the CMD, errors in T$_{\rm eff}$ and $\log\, g$ cannot be considered 
independent: e.g., an error in T$_{\rm eff}$ of about 100 K would also 
correspond to an uncertainty of about 0.25-0.30 dex in $\log\, g$. 
So the contributions from these two terms must first be summed up 
algebrically, and the result be summed in quadrature along with all other 
independent contributions to give the final error due to uncertainties in 
all the adopted parameters. These typically range from 0.06 dex for the 
brightest stars in the sample up to 0.10 dex for the faintest stars observed 
at the RHB level.

\begin{table} 
\caption{Sensitivities of $[$Na/Fe$]$ abundances to demonstrative errors in 
the atmospheric parameters} 
\begin{tabular}{lrrrr} 
\hline 
Element & $\Delta T_{\rm eff}$ & $\Delta \log g$ &$\Delta$[A/H] &$\Delta v_t$ \\ 
        & +100~K               & +0.3~dex  & 0.2~dex   & +0.2~km\, s$^{-1}$ \\ 
\hline 
\multicolumn{5}{c}{star 51871: T$_{\rm eff} = $3970 K, $\log\, g=0.62$ } \\ 
$[$Na/Fe$]$ & +0.205 & $-$0.105 & +0.011 &  $-$0.008 \\ 
\\ 
\multicolumn{5}{c}{star 39060: T$_{\rm eff} = $4357 K, $\log\, g=1.28$ } \\ 
$[$Na/Fe$]$ & +0.172 & $-$0.132 & +0.028 &  $-$0.016 \\ 
\\ 
\multicolumn{5}{c}{star 56136: T$_{\rm eff} = $4905 K, $\log\, g=2.38$ } \\ 
$[$Na/Fe$]$ & +0.292 & $-$0.156 & +0.041 &  $-$0.017 \\ 
\hline 
\end{tabular} 
\label{t:sensitivity} 
\end{table} 
 
To these, one has to add the uncertainties in the measurements. These random 
errors (including e.g. the continuum placement, the visual estimate of Na 
abundances from comparison with synthetic spectra) can be evaluated from 
the rms deviations of the average of Na abundances given by the two Na D 
lines. They range typically from 0.05 dex at the RGB-tip, to 0.06 dex around 
4350 K, and to 0.07 dex for the faintest stars in the sample. 
 
The resulting total random errors then vary from 0.08-0.09 dex for brighter 
stars and increase to as much as 0.12 dex at the RHB level. The random error 
bars are displayed in the upper panel of Fig.~\ref{f:nateff}. 
Hence, the observed spread in Na abundances is more than 5 times the error for 
the bright part of the sample and as much as 8-10 times the error when 
considering the lower luminosity giants. 
 
A first, firm result of these observations is then that large variations do 
exist in the abundances of elements produced in proton-capture reactions at all 
luminosities along the Red Giant Branch of NGC 2808, as shown by the large 
spread in [Na/Fe] abundances. 
 
\begin{figure} 
\psfig{figure=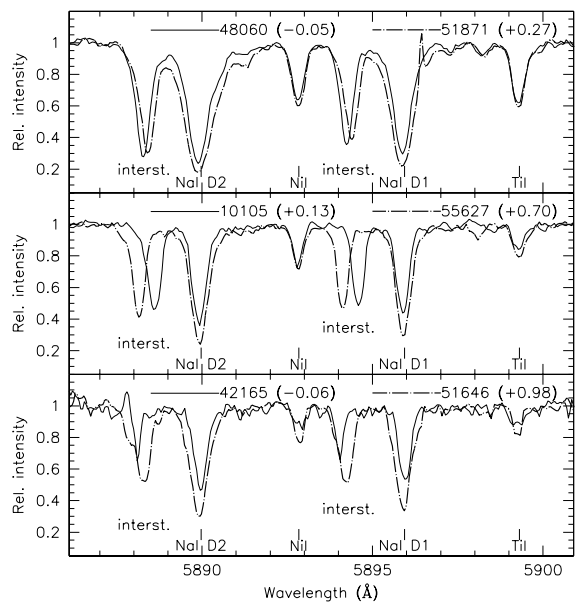,width=8.8cm,clip=} 
\caption[]{Observed spectra of pairs of stars with different strenghts in Na D 
lines, at the top (3970 K), in the middle (4450 K) and at the bottom (4906 K) 
of the temperature range spanned by our sample. Star identifications 
are given at the top of each panel, together with the [Na/Fe] value. 
A few photospheric lines are indicated, while 'interst.'
shows the position of the interstellar Na D lines.}
\label{f:comparana} 
\end{figure}

Apart from specific computations, Fig.~\ref{f:comparana} allows to immediately 
appreciate how much Na can differ between pairs of stars having the same 
temperature, hence the same evolutionary physical status. 
 
\section{Comparison with the Na distribution in other Globular Clusters} 
 
The proton capture elements as Na in RGB stars have been studied in several 
GC's, mainly by the Lick and Texas groups. We may compare here our findings 
with what has been derived for the following GC's ([Fe/H] values are all 
on the Carretta \& Gratton 1997 metallicity scale): 
M5 (Ivans et al. 2001; [Fe/H]=--1.11),
M13 (Pilachowski et al. 1996b, Kraft et al. 1997; [Fe/H]=--1.39), 
M15 and M92 (Sneden et al. 2000; [Fe/H]=--2.12 and --2.16, respectively). 
M4, studied by Ivans et al. (1999), will not be considered here 
because severe differential reddening makes the separation of 
true RGB from AGB stars very difficult. 
 
We have taken the [Na/Fe] values from the above quoted papers,  
together with the photometric information that comes originally from: 
Sandquist et al. (1996, M5), Cudworth \& Monet (1979, M13), Cudworth 
(1976, M15), and Rees (1992, M92). The distance 
moduli $(m-M)_V$ and reddenings were taken from Harris (1996). 
These [Na/Fe] values come from moderately high (R=11000) to very 
high (R=60000) resolution spectra. The analyses 
differ from ours in some parts: in some cases Na abundances have been 
derived from the 5682-88 \AA ~doublet (M5 and M13), in others from the Na D 
lines (M15 and M92); the Bell et al. (1976) grid of model atmosphere was 
usually used by the Texas-Lick group; finally, no 
correction for departure from LTE has been applied (this is slightly less 
important when the 5682-88 \AA ~lines are used). 
 
\subsection{NGC 2808 and M13} 
 
The largest samples of RGB stars are for M13 (112) 
and NGC 2808 (81), while less than about 30 RGB stars have been 
observed in the other GC's. The first comparison then will be done between 
these two GC's, since statistical significance and large sample of stars in 
different evolutionary phases are required to disentangle possible 
evolutionary effects, if any, overimposed to primordial variations. Moreover, 
Pilachowski et al. (1996b, hereafter PSKL) and Kraft et al. (1997) made a case 
for evolutionary effects being strongly at work in M13, which is also the 
template cluster as far as extreme oxygen depletion in RGB stars is concerned. 
 
The two clusters have been observed at similar resolutions (R $\sim$ 11000 
and 15000 respectively for M13 and NGC 2808), in both cases Na abundances come 
from spectral synthesis (5682-88 \AA ~lines, and Na D lines, respectively), 
the metallicities are quite similar (0.25 dex less for M13), and only true RGB 
stars are considered in the comparison. 
 
We note here that 13 more RGB tip stars were observed in NGC 2808, with UVES, 
and preliminary Na abundances were derived from abundance analysis based on 
the  EW's of four lines, the subordinate lines 5862-88~\AA\ and 6154-60~\AA. 
We do not include these stars in the present analysis to maintain a strict 
homogeneity in our database, however it's worth mentioning that they agree 
very well with the present results. 
 
To evaluate the (possible) presence of evolutionary effects we have to compare 
samples of stars in different phases along the RGB. PSKL define as "RGB tip" 
stars those having logg $\lsim$ 1, since they find an abrupt change in the Na 
abundance distribution at this point, with [Na/Fe] becoming larger and less 
dispersed for higher luminosity stars. We find a similar separation, but in our 
case the Na abundances decrease in the RGB tip stars. 
While it is not clear if this point has some physical relevance, e.g. marking 
the onset of a particular regime when the star approaches the last stages of 
lifetime as a red giant, in the following we will adopt logg = 1.02 as 
separator (as done e.g., in M5 by Ivans et al. 2001)  between RGB-tip and 
lower-RGB stars (see also PSKL for a discussion on this point). 
 
However, before attempting any comparison, we have to spend a few cautionary 
words on the various factors that may introduce spurious differences. 
When we compare results for bright and fainter objects, the larger random 
errors that are intrinsic to the analysis of fainter stars can produce an 
artificial increase in the abundance dispersion. 
Moreover, when we compare abundance analyses done by different groups we also 
have to take into account the different tools adopted: the use of the 
Kurucz or BEGN models for atmospheres, or the choice about correcting or not 
for NLTE, both introduce a tilt (see Fig.~\ref{f:nateff}) that influences 
the median values of the Na distributions. 
 
\begin{figure} 
\psfig{figure=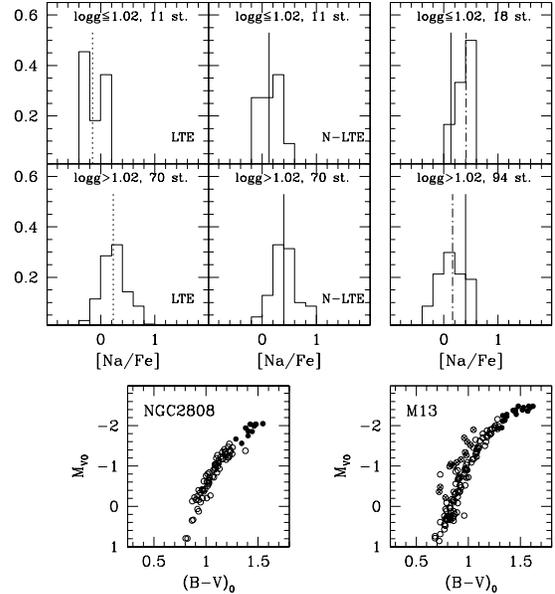,width=8.8cm,clip=}
\caption[]{ Bottom panels: CMD's for the RGB stars studied in NGC 2808 and M13; 
filled and open circles represent stars with $\log\, g \le$ and $ >1.02$, 
respectively; crosses indicate objects not considered in the histograms because 
they are possible AGB stars. 
Upper panels: histograms for the [Na/Fe] values, 
for $\log\, g \le$ and $>$ 1.02 respectively, normalized to the total number of 
stars used in the adopted range (labelled in the box). For NGC 2808 we plot
both the LTE and NLTE cases.
The vertical 
lines (dotted: NGC 2808 LTE; solid: NGC 2808 NLTE; dashed-dotted: M 13) 
indicate the median values for each considered range and case; the
value for the NLTE distributions in NGC 2808 are shown also in the M13
panels for comparison.}
\label{f:comphisto1} 
\end{figure} 
 
\begin{figure} 
\psfig{figure=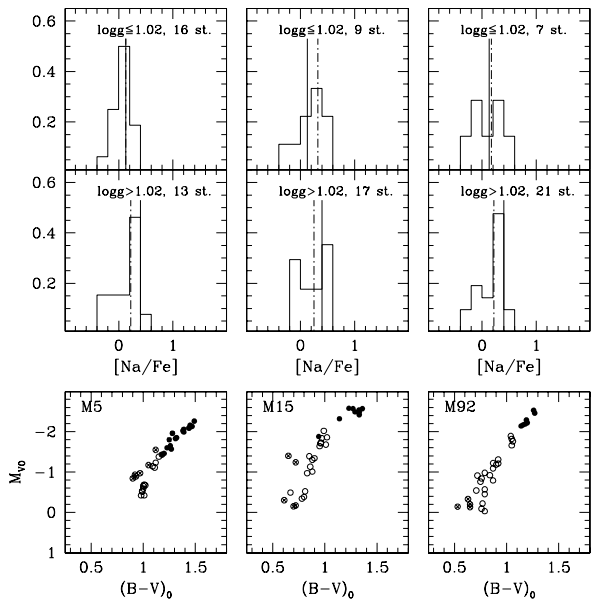,width=8.8cm,clip=} 
\caption[]{The same as Fig. 8, but for M5, M15, and M92.} 
\label{f:comphisto2} 
\end{figure} 
 
Fig.~\ref{f:comphisto1} (lower panels) shows the distribution along the RGB's 
of the observed stars for NGC 2808 and M13, in the dereddened (B-V)$_0$, $M_V$ 
plane, while in the upper panels are plotted the distributions of the [Na/Fe] 
ratios for RGB tip and lower RGB stars; the vertical lines indicate 
the median value of each distribution. 
NGC 2808 and M13 appear to behave quite differently in both bins/ranges. 
As already noted by PSKL, in M13 the brightest stars are skewed towards high Na 
abundances (the median of the [Na/Fe] values is +0.41 dex) and show a small 
dispersion, while among the lower RGB stars there is a number of objects 
that are more Na-poor (median +0.16 dex) and with a larger dispersion. 
Notice that the - small - differential corrections 
(needed to translate abundances derived with BEGN models to our own scale) 
between the two bins would only increase the difference, shifting  the 
medians apart. 
 
In NGC 2808 the opposite seems to be true: the RGB tip stars are Na-poorer, on 
average: a median value of 0.13 dex is found, to be compared with 0.40 dex for 
lower luminosity RGB stars. 
When running simple Kolmogorov-Smirnov tests, these differences (between the 
two bins for the same GC, and between GC's) appear significant: the two 
ranges in $\log\, g$ (or luminosity and/or temperature) have very low probability 
of being extracted from the same parent population. 
Moreover, both ranges seem to be statistically different between M13 and 
NGC 2808. 
While a larger spread at hotter temperatures is expected because of 
the worsening of spectra quality at lower luminosities, it
is the median value that matters, as far as the interpretation of this data is
concerned.
 
The interpretation by PSKL of this pattern of abundances in M13 was that, 
in addition to the primordial variations among stars, an evolutionary effect 
is present, that increases the Na abundance somewhat differently for each 
star, since the mixing efficiency is different from stars to star.  
This evolutionary Na enhancement is dominant in the upper RGB, as
witnessed also by the lowering of carbon isotopic ratios and carbon abundances
(accompanied by increasing nitrogen abundances climbing up the RGB in several
clusters, see the review by Kraft 1994 for references).
Instead, in NGC 2808 this enhancement of the number of Na-rich stars is 
not seen, and the variation goes the opposite way, as in other clusters 
(see below). 
 
The conclusions of this comparison are that: 
 
\noindent 
i) in NGC 2808 the evolutionary effect claimed in M 13- if at all present
- is only at the  noise level. This would imply that the Na abundance
distribution along the RGB  is most probably of primordial origin, either
because it was imprinted in  the gas from which these stars originated, or
because it was produced by  an early pollution event due to a first generation
of intermediate-mass  AGB stars; 
 
\noindent 
ii) this finding seems to be quite at odds with what one may expect from 
the distribution of stars on the HB, which is rather peculiar in NGC 2808 
(an important example of ``second-parameter'' effect). 
Gratton (1982) already noted that about one 
fourth of the HB stars lie on the blue HB, which is quite unexpected given the 
metallicity of the cluster. 

On the working hypothesis (recently explored 
by D'Antona et al. 2002) that Na-rich stars are also enriched in Helium 
(as well as depleted in Oxygen, 
as found in NGC 2808 by Carretta et al. 2003b), 
they would populate exactly the blue part of the HB.
This, however,  
would require that about 1/4 of the stars in the upper 
RGB stand out as Na-rich objects, which is maybe observed in M13 but not 
very clearly in NGC 2808. 
The scenario proposed by D'Antona et al. (2002) also implies that more than
one episode of star formation may have occurred within a GC, producing a 
bi(multi)-modal mass distribution on the HB. A similar type of distribution
in the Na abundances should then be present along the RGB, which however is
not observed with a sufficient degree of confidence.

In our opinion, the lack of a clear peak at high Na values in the
distribution for NGC 2808 is an indication that the
possible link between HB morphology (extension and mass distribution) and other
parameters such as the extent of chemical anomalies may be not very tight.
 
\subsection{Comparison with other clusters: M5, M15, M92} 
 
As a result of the above comparison, can we deduce that NGC 2808 is a 
peculiar GC? Or is instead M13 that behaves unusually? 
Already Ivans et al. (2001) in their study on M5 noted that this cluster 
does not behave like M13 (see their fig. 11). 
To M5 we may add M15 and M92: even with smaller 
numbers, the samples in these three GCs reach down to the same luminosity 
level (M$_V \sim$ 0) of the bulk of our program stars, so  we feel entitled 
to repeat the comparison. 
 
The results are displayed in Fig.~\ref{f:comphisto2}. 
In all three cases the [Na/Fe] distribution does not vary  significantly 
(as deduced again by the Kolmogorov-Smirnov test) between the two bins (upper 
and lower RGB), and the median values are quite similar, in particular for the 
lower RGB.  
Again, based on this test, M13 appears to stand out among globular clusters, 
since none of the examined clusters shows a clearcut shift toward Na-rich 
stars when climbing up the RGB. This is 
not surprising, as M13 is peculiar in other ways, like its anomalously 
extendend Na-O anticorrelation. It seems that in every other cluster, the
scatter introduced by possible evolutionary effects is very small and does not
affect sensibly the resulting abundances. This is more evident in the 3 other
clusters shown in Fig.~\ref{f:comphisto2}, where stars are observed
preferentially on the upper part of the RGB. An increase of the scatter is
hardly detectable.

\section{SUMMARY} 
 
We have analysed moderately high-resolution spectra of more than 80 red giant 
stars in the globular cluster NGC 2808, taken during the Science Verification 
program of the FLAMES multi-object spectrograph at the VLT. 
Atmospheric parameters were derived from the photometric information, radial 
velocities (which permitted to exclude a few field interlopers, leaving a 
sample of 81 RGB stars) were derived from the spectra. 
We examined in this paper only the Sodium abundances, that were derived from 
the Na {\sc i} D lines and spectral synthesis, and corrected for NLTE effects. 
The main results of our analysis are: 
 
1- There are large variations in Na abundances at all luminosity levels along 
the RGB in NGC 2808. This is an intrinsic spread, since it is much larger than 
possible uncertainties in the derived abundances (being from 0.5 to 1 dex, to 
compare with errors of about 0.08 dex increasing to 0.12 dex at the very faint 
end of our sample). 
While this is not surprising, being commonly observed in several globular 
clusters, this is the first time that detailed abundance analysis uncovers this 
pattern in NGC 2808. 
This spread is not explained by standard stellar evolutionary models, and 
its presence requires variable amount of yields from 
proton capture fusion on Neon, that can only happen in the interior region 
near the H-burning shell, in the same star examined or in other objects that 
were then able to somehow pollute its surface. 
 
2- We have compared RGB-tip to lower-RGB stars to see whether there is any 
evolutionary effect acting like in M13, and enhancing Na for the brighter, 
more evolved giants. This effect does not appear to be important in NGC 2808, 
since brighter RGB stars have lower average Na abundances; if a similar 
mechanism is at work in this cluster, it only contributes adding some noise 
to a primordial dispersion. Our data are not sufficient to distinguish if the 
different composition was pre-existing the star forming gas, or is due to  
pollution on the star surface. 
 
3- This Na abundance dispersion, and the (absence of) variation towards 
the RGB tip is similar to what was found for other GCs, e.g. M5, M15, and M92, 
but at odds with what is seen in M13, the only other cluster for which a 
very large sample of RGB stars has been observed. 
 
Homogeneous analysis of large samples of stars observed at high resolution 
and high S/N is now possible for all Galactic globular clusters, with the 
new multiobject spectrographs mounted on  8-10m class telescopes. This 
is the road we have to follow if we wish to make significative improvements on 
our knowledge of stellar formation and evolution in globular clusters.

\begin{acknowledgements} 
{ This research has made use of data taken during FLAMES Science Verification. 
We wish to warmly thank Luca Pasquini for having co-ordinated the building of 
such a powerful instrument, and the ESO staff at Paranal for their excellent 
work during SV. 
We are specially indebted to Raffaele Gratton for many interesting and helpful 
suggestions and discussions that greatly improved the present work. We also 
thank the referee (Bob Rood) for his suggestions.} 
\end{acknowledgements}

\end{document}